\documentstyle[aps,preprint]{revtex}
\widetext
\draft
\tighten

\begin{document}

\preprint{EFUAZ FT-95-12}

\title{Significance of the spinorial basis\linebreak in relativistic
quantum mechanics\thanks{Submitted
to {\it ``American Journal of Physics".}}}

\author{{\bf Valeri V. Dvoeglazov}\thanks{On leave of absence from
{\it Dept. Theor. \& Nucl. Phys., Saratov State University,
Astrakhanskaya ul., 83, Saratov\, RUSSIA.} Internet
address: dvoeglazov@main1.jinr.dubna.su}}

\address{
Escuela de F\'{\i}sica, Universidad Aut\'onoma de Zacatecas \\
Antonio Doval\'{\i} Jaime\, s/n, Zacatecas 98000, ZAC., M\'exico\\
Internet address:  VALERI@CANTERA.REDUAZ.MX}

\date{March 16, 1995}
\maketitle

\begin{abstract}
The problems connected with a choice of the spinorial basis
in the $(j,0)\oplus (0,j)$ representation space are discussed.
It is shown to have profound  significance in relativistic quantum theory.
{}From the methodological viewpoint this
fact is related with the important dynamical role played by space-time
symmetries for all kind of interactions.
\end{abstract}


\newpage

We have become accustomed to thinking of the particle world
from a viewpoint of the principle of gauge invariance. Profound
significance of this principle seems to be clear for everybody
and it deserves to be in the place that it occupies now.
Remarkable experimental confirmations of both quantum
electrodynamics~\cite{DFT} and its non-Abelian extensions
(Weinberg-Salam-Glashow model, quantum chromodynamics),
ref.~\cite{Rosner,QCD}, proved its applicability.
Nevertheless, let us still not forget that firstly the principle
has been deduced from the interaction of charge particles
with electromagnetic potential. At the same time,
it has been long recognized that for other kind of particles
(namely, for truly neutral particles that are supposed to be described by
self/anti-self charge conjugate states) a change of phase leads to
destroying  self/anti-self conjugacy~\cite{Furry}. It is in this field
of modern science (neutrino physics, gluon contributions in QCD etc.)
that we have now most consistent indications for new physics. Without any
intention to shadow great achievements~\cite{Dirac} of the theories based
on the use of 4-vector potentials I am going to look into the
subject from a little bit different point of view. I would like to discuss
here the constructs based on the use of the $(j,0)\oplus (0,j)$
Lorentz group representations for description of the particle
world and interactions in it.
I hope that presented thoughts could be useful for deeper
understanding surprising symmetries of the Dirac equation and an
unexpectable rich structure of the $(j,0)\oplus (0,j)$ representation
space\footnote{Of course, spin-1/2 fermions, that transform on the $(1/2,0)
\oplus (0,1/2)$ representation of the Lorentz group, could be
considered as particular cases. The discussion of recent  achievements
in development of the Weinberg $2(2j+1)$ component theory~\cite{Weinberg}
could be found in ref.~\cite{DVO-old}.} by students and, perhaps, by higher
school professors.

The spinorial basis  in the standard representation of the Dirac
equation\footnote{I use here and below a notation of
refs.~\cite{Ryder,DVA0,DVAG}.
For the 4-momentum of a particle in the rest
one uses $\overcirc{p}^\mu$.}:
\begin{eqnarray}
u^{(1)} (\overcirc{p}^\mu) = \sqrt{m}\pmatrix{1\cr 0\cr 0\cr 0\cr},\quad
u^{(2)} (\overcirc{p}^\mu) = \sqrt{m}\pmatrix{0\cr 1\cr 0\cr 0\cr},\quad
v^{(1)} (\overcirc{p}^\mu) = \sqrt{m}\pmatrix{0\cr 0\cr 1\cr 0\cr},\quad
v^{(2)} (\overcirc{p}^\mu) = \sqrt{m}\pmatrix{0\cr 0\cr 0\cr 1\cr}
\end{eqnarray}
is well understood and acceptable by everybody for description
of a Dirac particle. However, let ask ourselves, what forced
us to choose it?.. Let me attack the problem of the choice of
a spinorial basis from the most general position.

\smallskip

I am going to consider theories based on the following four
postulates:
\begin{itemize}

\item
For arbitrary $j$ the right $(j,0)$ and the left $(0,j)$ handed spinors
transform in the following ways (according to the Wigner's
ideas~\cite{Wigner2,Wigner}):
\begin{mathletters}
\begin{eqnarray}
\phi_R (p^\mu)\, &=& \,\Lambda_R (p^\mu \leftarrow \overcirc{p}^\mu)\,\phi_R
(\overcirc{p}^\mu) \, = \, exp (+\,{\bf J} \cdot {\bbox \varphi})
\,\phi_R (\overcirc{p}^\mu)\quad,\\
\phi_L (p^\mu)\, &=&\, \Lambda_L (p^\mu \leftarrow \overcirc{p}^\mu)\,\phi_L
(\overcirc{p}^\mu) \, = \, exp (-\,{\bf J} \cdot {\bbox \varphi})\,\phi_L
(\overcirc{p}^\mu)\quad.\label{boost0}
\end{eqnarray}
\end{mathletters}
$\Lambda_{R,L}$ are the matrices for Lorentz boosts; ${\bf J}$ are
the spin matrices for spin $j$; ${\bbox \varphi}$ are parameters of the
given boost.  If restrict ourselves by the case of bradyons they are
defined, {\it e.~g.}, refs.~\cite{Ryder,DVA0}, by means of:
\begin{equation}\label{boost}
\cosh (\varphi) =\gamma = \frac{1}{\sqrt{1-v^2}} = \frac{E}{m},\quad
\sinh (\varphi) = v\gamma = \frac{\vert {\bf p}\vert}{m},\quad
\hat {\bbox \varphi} = {\bf n} = \frac{{\bf p}}{\vert {\bf p}\vert}\quad.
\end{equation}

\item
$\phi_L$ and $\phi_R$ are the eigenspinors of the helicity
operator $({\bf J}\cdot {\bf n})$:
\begin{equation}
({\bf J}\cdot {\bf n})\,\phi_{R,L} \,=\, h \,\phi_{R,L}
\end{equation}
($h = -j, -j+1,\ldots j$  is the helicity quantum number).

\item
The relativistic dispersion relation
$E^2 -{\bf p}^2 =m^2$ is hold for free particles.

\item
Physical results do not depend on rotations of spatial
coordinate axes (in other words: the 3-space is uniform).

\end{itemize}

\smallskip

\noindent
Since spin-1/2 particles are most important in physical
applications and, moreover, the Maxwell's spin-1 equations
can be written in the similar 4-component form, {\it e.g.},
ref.~\cite{Sachs}, let me concentrate in the
analysis of the $(1/2,0)\oplus
(0,1/2)$ representation space. For the sake of compact description
let denote 2-spinors (left- or right-handed) as $\xi$.
{}From the condition (see the second item):
\begin{equation}
{1\over 2} \,({\bbox \sigma}\cdot {\bf n})\, \xi \,
=\, \pm\, {1\over 2}\, \xi
\end{equation}
and by using the expressions for
${\bf n}$ in spherical coordinates:
\begin{mathletters}
\begin{eqnarray}
n_x &=& \sin \theta \cos\phi\quad,\\
n_y &=& \sin\theta \sin\phi\quad,\\
n_z &=& \cos \theta\quad,
\end{eqnarray}
\end{mathletters}
we find that the Pauli spinor $\xi=column (\xi_1\quad \xi_2)$
answering for the eigenvalue $h=1/2$ of the helicity operator can be
parametrized
as
\begin{equation}
\xi_{+\,1/2} \,=\, \pmatrix{\xi_1 \cr
\tan \,(\theta /2)\, e^{i\phi} \,\xi_1} \quad\mbox{or}
\quad \xi_{+\,1/2} \,=\, \pmatrix{\cot \,(\theta /2)\, e^{-i\phi}\,\xi_2\cr
\xi_2}
\end{equation}
in terms of the azimutal $\theta$ and the polar $\phi$ angles associated
with the  vector ${\bf p} \rightarrow 0$, refs.~\cite{Itzykson,DVAG}.
The one, answering for the $h=-1/2$ eigenvalue, as
\begin{equation}
\xi_{-\,1/2} \,=\, \pmatrix{\xi_1 \cr
- \cot\, (\theta /2)\, e^{i\phi}\, \xi_1} \quad \mbox{or}
\quad \xi_{-\,1/2} \,=\, \pmatrix{- \tan \,(\theta /2)\, e^{-i\phi}\,\xi_2\cr
\xi_2}\quad.
\end{equation}
{}From the normalization condition $\xi^\dagger_{\pm 1/2} \xi_{\pm 1/2} =N^2$
(with $N^2$ being a normalization factor)
we have that the form of spinors can be chosen\footnote{The second
parametrization differs from the first one by an overall
phase factor, what does not have influence on physical results.}
\begin{mathletters}
\begin{eqnarray}
\xi_{+\,1/2} \,&=&\,N\,e^{i\,\vartheta^{+}_1}\,
\pmatrix{\cos\,(\theta/2)\cr
\sin\,(\theta/2)\,e^{i\,\phi}}\quad \mbox{or}\quad
\xi_{+\,1/2} \,=\,N\,e^{i\,\vartheta^{+}_2}\,
\pmatrix{\cos\,(\theta/2)\,e^{-i\,\phi}\cr
\sin\,(\theta/2)\cr}\quad,\label{exphia}\\
\xi_{-\,1/2}\,&=&\,N\,e^{i\,\vartheta^{-}_1}\,
\pmatrix{\sin\,(\theta/2)\cr
-\,\cos\,(\theta/2)\,e^{i\,\phi}}\quad\mbox{or}\quad
\xi_{-\,1/2}\,=\,N\,e^{i\,\vartheta^{-}_2}\,
\pmatrix{-\,\sin\,(\theta/2) \,e^{-i\,\phi}\cr
\cos\,(\theta/2)\cr}\quad.\label{exphib}
\end{eqnarray}
\end{mathletters}
This parametrization coincides with Eqs. (22a,22b) of ref~[12b]
and with the formulas of ref.~\cite[p.87]{Itzykson} within definitions
of overall phase factors $\vartheta^{\pm}$. Let me note
useful identities:
\begin{equation}
\xi_{+\,1/2} (\overcirc{p}^{\prime\,\,\mu})\,=\, e^{i(\vartheta^+ -
\vartheta^-)}\xi_{-\,1/2} (\overcirc{p}^\mu)\quad, \quad
\xi_{-\,1/2} (\overcirc{p}^{\prime\,\,\mu})\,=\, e^{i(\vartheta^- -
\vartheta^+)}\xi_{+\,1/2} (\overcirc{p}^\mu)\quad,
\end{equation}
where $\overcirc{p}^{\prime\,\,\mu}$ is the parity conjugated
4-momentum in the rest ($\theta^\prime = \pi-\theta$,
$\phi^\prime=\phi+\pi$).

If we know spin matrices for arbitrary $j$ one
could find similar parametrizations for spinors of higher dimensions
by resolving the set of equations of the $(2j+1)$- order for each value of
the helicity\footnote{See, {\it e.~g.}, the formulas (23a-c)
in ref.~[12b] and below.}. Let me note that one has a certain
freedom in a choice of the spinorial basis in the $(1/2,0)$ (or $(0,1/2)$)
space since, according to the fourth postulate, physical results do not
depend on rotations of the spatial axes and, furthermore, one has arbitrary
phase factors $e^{i\vartheta^{\pm}}$ .  Therefore, the common-used
choice (${\bf p} \,\vert \, \vert OZ$)
\begin{eqnarray}
\xi_{+\, 1/2} = \pmatrix{1\cr 0\cr},\quad
\xi_{-1/2} = \pmatrix{0\cr 1\cr}\quad.
\end{eqnarray}
is only a convenience.

In the Dirac equation one has two kind of spinors ($\phi_R$ and $\phi_L$).
In refs.~\cite{Ryder,DVA0,DVAG} the relation between them in the rest
frame
\begin{equation}
\phi_R (\overcirc{p}^\mu) =\pm \phi_L (\overcirc{p}^\mu)
\label{RB}
\end{equation}
has been named as the Ryder-Burgard relation (see also~\cite{Novozh}).
It was shown (see footnote \# 1 in~[12b])
that the relation (\ref{RB}) can be used to derive the Dirac
equation, the equation that describes eigenstates of the charge operator.
Moreover, if accept this form of the relation for $(1,0)\oplus (0,1)$
bispinors one can construct an example
of the Foldy-Nigam-Bargmann-Wightman-Wigner (FNBWW) type
quantum field theory~\cite{Nigam,Wigner2,DVA0}.  The remarkable feature
of this {\it Dirac}-like modification of the Weinberg
theory~\cite{Weinberg} is the fact that boson and its antiboson have
opposite relative intrinsic parities (like the Dirac fermion).

However, nobody forbids us to take more general form of Eq. (\ref{RB}).
Let assume that $\phi_R (\overcirc{p}^\mu)$
and $\phi_L(\overcirc{p}^\mu)$
are connected by an arbitrary linear transformation
with the complex matrix ${\cal A}$, namely, $\phi_R (\overcirc{p}^\mu) =
{\cal A} \phi_L (\overcirc{p}^\mu)$. The unit matrix and Pauli three
$\sigma$-matrices form a complete set. Therefore, the matrix
corresponding to the linear transformation ${\cal A}$ can be expanded
in this complete set with the complex
coefficients $c_i$:\footnote{The signs $\pm$ should be referred
to the helicity of the spinors.}
\begin{eqnarray}
\lefteqn{\phi_R^\pm (\overcirc{p}^\mu)
= {\cal A} \phi_L^\pm (\overcirc{p}^\mu) =
\left [\openone \,\, c_1^0+ {\bbox \sigma}\cdot{\bf c}_1\right ]
\phi_L^\pm (\overcirc{p}^\mu)
=\nonumber}\\
&=& \left [ c^0_1 \pm (\vert {\Re}e\, {\bf c}_1\vert + i\,\vert
{\Im}m\,{\bf c}_1\vert) \right ] \phi_L^\pm
(\overcirc{p}^\mu)
= e^{i\alpha_{\pm}} \,\phi_L^\pm (\overcirc{p}^\mu)\quad.
\end{eqnarray}
Above we have used that $\phi_L$ and $\phi_R$ are the eigenspinors of
the helicity operator and have chosen the parametrization
of the coefficients
$\left [c^0_1 \pm (\vert {\Re}e\, {\bf c}_1\vert +
i\vert {\Im}m\, {\bf c}_1 \vert ) \right ] = e^{i\alpha_\pm}$.
The modulus of the bracketed quantity
(the determinant of the ${\cal A}$ matrix) should
be equal to the unit from the condition of invariance of
the norm of spinors. The equation (\ref{RB}) answers for
the particular choices of $\alpha_\pm = 0,\, \pm \pi$.
By using the generalized Ryder-Burgard relation
and the fact that
\begin{equation}
\left [\Lambda_{L,R} (p^\mu \leftarrow \overcirc{p}^\mu)\right ]^{-1}
= \left [\Lambda_{R,L} (p^\mu \leftarrow
\overcirc{p}^\mu)\right ]^\dagger\quad,
\end{equation}
we immediately obtain the ``generalized" Dirac
equation:\footnote{Please, do not forget that
the Lorentz boost matrices are Hermitian for a finite
representation of the group.}
\begin{mathletters}
\begin{eqnarray}
\phi_R^\pm (p^\mu) \,&=&\, \Lambda_R (p^\mu \leftarrow \overcirc{p}^\mu)\,
\phi_R^\pm (\overcirc{p}^\mu) = \,e^{i\alpha_{\pm}}
\,\Lambda_R  (p^\mu \leftarrow \overcirc{p}^\mu)
\,\phi_L^\pm (\overcirc{p}^\mu) =\nonumber            \\
&=&\, e^{i\alpha_{\pm}}\,\Lambda_R (p^\mu
\leftarrow \overcirc{p}^\mu) \,\Lambda_L^{-1} (p^\mu \leftarrow
\overcirc{p}^\mu)\,\phi_L^\pm (p^\mu)\quad, \label{eq1} \\
&&\nonumber\\
\phi_L^\pm (p^\mu) \,&=&\, \Lambda_L (p^\mu \leftarrow \overcirc{p}^\mu)
\,\phi_L^\pm (\overcirc{p}^\mu) \, = \, e^{-i\alpha_{\pm}}\,
\Lambda_L (p^\mu \leftarrow \overcirc{p}^\mu) \, \phi_R^\pm
(\overcirc{p}^\mu) =\nonumber\\
&=& \,e^{-i\alpha_{\pm}} \,\Lambda_L (p^\mu
\leftarrow \overcirc{p}^\mu) \,\Lambda_R^{-1} (p^\mu \leftarrow
\overcirc{p}^\mu) \,\phi_R^\pm (p^\mu)\quad.\label{eq2}
\end{eqnarray}
\end{mathletters}
By using definitions of the Lorentz boost (2,\ref{boost})
one can re-write the equations (\ref{eq1},\ref{eq2}) in the matrix form
(provided that $m\neq 0$):
\begin{eqnarray}
\pmatrix{-m e^{-i\alpha_{\pm}}& p_0 +({\bbox \sigma}\cdot{\bf p})\cr
p_0 - ({\bbox \sigma}\cdot{\bf p}) & -m e^{i\alpha_{\pm}}\cr}
\pmatrix{\phi_R (p^\mu) \cr \phi_L (p^\mu)} = 0\quad,
\end{eqnarray}
or
\begin{equation}
\left ( \hat p -m{\cal T} \right ) \Psi (p^\mu) = 0\quad,
\end{equation}
with
\begin{eqnarray}
{\cal T} = \pmatrix{e^{-i\alpha_{\pm}}& 0\cr 0 &
e^{i\alpha_{\pm}}\cr}\quad.
\end{eqnarray}

Let us note the particular cases:
\begin{mathletters}
\begin{eqnarray}
\alpha_\pm &=& 0, 2\pi \quad : \quad (\hat p - m)\Psi
=0\quad, \label{DE1}\\
\alpha_\pm &=& \pm\,\, \pi \quad\,\,\, : \quad (\hat p + m)\Psi =0\quad,
\label{DE2}\\
\alpha_\pm &=& +\pi/2 \quad : \quad (\hat p + im\gamma_5)\Psi
=0\quad,\label{DDE1}\\
\alpha_\pm &=& -\pi/2 \quad : \quad (\hat p - im\gamma_5)\Psi =0 \quad.
\label{DDE2}
\end{eqnarray}
\end{mathletters}
The equations (\ref{DE1},\ref{DE2}) are the well-known Dirac equations for
positive- and negative-energy bispinors in the momentum space.
The equations of the type (\ref{DDE1},\ref{DDE2}) had also
been disussed in the old literature, {\it e.~g.},~\cite{Sokolik}.
They have been named as the
Dirac equations for 4-spinors of the second kind~\cite{Cartan,Gelfand}.
Their possible relevance to description of neutrino has been mentioned
in the cited papers.  Let still note, this idea has been proposed before an
appearance of the two-component model of Landau, Lee, Salam and Yang.

Next, since spinors are, in general, the complex quantities
it is possible to set up the Ryder-Burgard relation in the following form:
$\phi_R (\overcirc{p}^\mu) = {\cal B} \phi_L^* (\overcirc{p}^\mu)$.
Let me remind that the operation of complex conjugation is
not linear operator. Therefore, these two forms of the relation
are not equivalent.
It is more convenient to expand the ${\cal B}$ in the other complete
set:\footnote{We know that after a multiplication by a non-singular
matrix the property of being a complete set  is hold.} $\sigma_2$ and
$\sigma_i \sigma_2$.  We want again that the norm
is conserved:\, $\phi_{R,L}^\dagger \phi_{R,L} =
\phi_{R,L}^T \phi_{R,L}^* = N^2$.
By using the procedure analogous to the above we come
to another form of the Ryder-Burgard relation:
\begin{eqnarray}
\lefteqn{\phi_R^\pm (\overcirc{p}^\mu) = {\cal B} [\phi_L^\pm
(\overcirc{p}^\mu)]^* = \left [ c^0_2 \, \sigma_2 + ({\bbox \sigma}\cdot {\bf
c}_2) \sigma_2\right ] [\phi_L^\pm (\overcirc{p}^\mu)]^* =\nonumber}
\label{rbc} \\
&=& \left [i\,c^0_2 \, \Theta_{[1/2]} \mp i\, (\vert {\Re}e\, {\bf c}_2\vert
+ i \, \vert {\Im}m\, {\bf c}_2 \vert )
\,\Theta_{[1/2]}\right ] [\phi_L^\pm (\overcirc{p}^\mu)]^*
= i\,e^{i\beta_{\mp}} \Theta_{[1/2]}\, [\phi_L^\pm
(\overcirc{p})]^*
\end{eqnarray}
and, hence, to the inverse one
\begin{equation}
\phi_L^\pm (\overcirc{p}^\mu) = - i\,e^{i\beta_{\mp}} \Theta_{[1/2]}
[\phi_R^\pm (\overcirc{p}^\mu)]^*\quad.\label{rbinv}
\end{equation}
We have used above that
$\sigma_2$ matrix is connected with the Wigner operator
$\Theta_{[1/2]}=-i\sigma_2$
and the property of the Wigner operator for any spin
$\Theta_{[j]} {\bf J}\Theta_{[j]}^{-1} = -{\bf J}^*$.
So, if $\phi_{L,R}$ is an eigenstate of the helicity
operator, then $\Theta_{[j]}\phi_{L,R}^*$ is the eigenstate with
the opposite helicity quantum number:
\begin{equation}
({\bf J}\cdot {\bf n}) \,\Theta_{[j]} \left [ \phi_{L,R}^h (p^\mu)\right ]^*
= -\, h \,\Theta_{[j]} \left [\phi_{L,R}^h\right ]^*\quad.
\end{equation}
Therefore, from Eqs. (\ref{rbc},\ref{rbinv}) we have
\begin{mathletters}
\begin{eqnarray}
\phi_R^\pm (p^\mu) &=& +\,i\,e^{i\beta_{\mp}}\, \Lambda_R (p^\mu \leftarrow
\overcirc{p}^\mu)\, \Theta_{[1/2]}\, [\Lambda_L^{-1}
(p^\mu \leftarrow \overcirc{p}^\mu)]^* \,
[\phi_L^\pm (p^\mu)]^*\quad,\label{cc1}\\
\phi_L^\pm (p^\mu) &=& -\,i\,e^{i\beta_{\mp}} \,\Lambda_L (p^\mu \leftarrow
\overcirc{p}^\mu) \,\Theta_{[1/2]}\, [\Lambda_R^{-1}
(p^\mu \leftarrow \overcirc{p}^\mu)]^* \,
[\phi_R^\pm (p^\mu)]^*\quad.\label{cc2}
\end{eqnarray}
\end{mathletters}
By using the mentioned property of the Wigner operator
we transform  Eqs. (\ref{cc1},\ref{cc2}) to
\begin{mathletters}
\begin{eqnarray}
\phi_R^\pm (p^\mu) &=& +\,ie^{i\beta_{\mp}} \Theta_{[1/2]}
[\phi_L^\pm (p^\mu)]^*\quad,\\
\phi_L^\pm (p^\mu) &=& -\,ie^{i\beta_{\mp}} \Theta_{[1/2]}
[\phi_R^\pm (p^\mu)]^*\quad.
\end{eqnarray}
\end{mathletters}
In the matrix form one has
\begin{equation}
\pmatrix{\phi_R (p^\mu)\cr \phi_L (p^\mu)} =
e^{i\beta_{\mp}} \pmatrix{0& i\Theta_{[1/2]}\cr
-i\Theta_{[1/2]} &0\cr}\pmatrix{\phi_R^* (p^\mu)\cr
\phi_L^* (p^\mu)} = S^c_{[1/2]} \pmatrix{\phi_R (p^\mu)\cr
\phi_L (p^\mu)}\quad,
\end{equation}
with $S^c_{[1/2]}$ being the operator of charge conjugation in
the $(1/2,0)\oplus (0,1/2)$ representation space, {\it e.~g.}~\cite{Ramond}.
In fact, we obtain conditions of self/anti-self charge congugacy:
\begin{equation}
\Psi (p^\mu) = \pm \Psi^c (p^\mu)\quad.
\end{equation}

Thus, depending on relations between left- and right-handed
spinors (in fact, depending on the choice of the spinorial basis) we
obtain physical excitations of the different physical nature.
In the first version of the Ryder-Burgard relation
we have the Dirac equations; in the framework of
the second version, neutral fermions\footnote{In fact, the
second definition of the Ryder-Burgard relation leads to the
Majorana-McLennan-Case spinors~\cite{Majorana,Case}. Recent
discussions of this construct could be found in refs.~\cite{DVA,DVAG,DV95}.}.

At last, the most general form of the Ryder-Burgard relation
is\,\footnote{We deal above with
the spinors of same helicities.
The student can reveal without any troubles, what happens
if we were connect the spinors of different helicities,
$\phi_R^\pm = {\cal A} \phi_L^\mp$  or $\phi_R^\pm =
{\cal B} [\phi_L^\mp ]^*$.}
\begin{equation}
\phi_R (\overcirc{p}^\mu) = {\cal A} \phi_L (\overcirc{p}^\mu) +
{\cal B} \phi_L^* (\overcirc{p}^\mu)\quad,
\end{equation}
what results in
\begin{mathletters}
\begin{eqnarray}
\phi_R (\overcirc{p}^\mu) \,&=&\,  A \,e^{i\alpha_{\pm}} \,\phi_L
(\overcirc{p}^\mu) + i\,B \,e^{i\beta_{\mp}}\, \Theta_{[1/2]}
\,\phi_L^* (\overcirc{p}^\mu)\quad,\\
\phi_L (\overcirc{p}^\mu) \,&=&\, A \,e^{-i\alpha_{\pm}} \,\phi_R
(\overcirc{p}^\mu) - i\,B \,e^{i\beta_{\mp}} \,\Theta_{[1/2]}
\,\phi_R^* (\overcirc{p}^\mu)\quad.
\end{eqnarray}
\end{mathletters}
The equation, that could be considered as
a mathematical generalization of the Dirac equation, is then
\begin{eqnarray}
\pmatrix{-1 & A\, e^{i\alpha_{\pm}}
\,\Lambda_R \,\Lambda_L^{-1} + i\,B \,e^{i\beta_{\mp}} \,\Theta_{[1/2]}\,
{\cal K}\cr
A \,e^{-i\alpha_{\pm}}\, \Lambda_L \,\Lambda_R^{-1} - i\,B \,e^{i\beta_{\mp}}
\,\Theta_{[1/2]} \,{\cal K} & -1\cr}
\pmatrix{\phi_R (p^\mu)\cr\phi_L (p^\mu)\cr} =0\quad,
\end{eqnarray}
where $A^2 +B^2 =1$ and ${\cal K}$ is the operation of complex conjugacy.
In a symbolic form it is re-written to
\begin{equation}
\left [A \,{\hat p \over m} + B \,{\cal T} \,S^c_{[1/2]}
- {\cal T}\right ]\Psi (p^\mu) = 0\quad.
\end{equation}
By using the computer algebra system MATEMATICA 2.2 it is easy
to check that the equation has the correct relativistic dispersion
(see the third item of the set of postulates).
What physical sense could be attached to this equation?

Let me now regard the problem of a choice of spinorial basis
in a $j=1$ case. In the presented consideration I use the Weinberg
$2(2j+1)$ component formalism~\cite{Weinberg}.
It is easy to show, by using the same procedure, that $j=1$
spinors $\xi$ can be parametrized, {\it e.~g.}, in the following form
\begin{mathletters}
\begin{eqnarray}
\xi_{+1} =
N\,e^{i\vartheta_+}\,\pmatrix{{1\over 2} (1+\cos\theta) e^{-i\phi}\cr
\sqrt{{1\over 2}} \sin\theta\cr
{1\over 2} (1-\cos\theta) e^{+i\phi}\cr}&&\quad , \quad
\xi_{-1} = N\,e^{i\vartheta_-}\,\pmatrix{-{1\over 2} (1-\cos\theta)
e^{-i\phi}\cr
\sqrt{{1\over 2}} \sin\theta\cr
-{1\over 2} (1+\cos\theta) e^{+i\phi}\cr}\quad,\\
&&\nonumber\\
\xi_{0} &=&  N\,e^{i\vartheta_0}\,
\pmatrix{-\sqrt{{1\over 2}}\sin \theta \,e^{-i\phi}\cr
\cos\theta\cr
\sqrt{{1\over 2}}\sin \theta \,e^{+i\phi}\cr}\quad,
\end{eqnarray}
\end{mathletters}
provided that they are eigenspinors of the helicity operator.
In the isotropic-basis representation the $j=1$ spin operators are
expressed, ref.~\cite{Var},\footnote{Of course, it is possible to choose
the so-called `orthogonal' basis
$({\bf J}_i)_{jk} = -i\epsilon_{ijk}$ because they are connected by
the unitary matrix ${\bf J}^{isotr.} = {\cal U} {\bf J}^{orth.} {\cal
U}^{-1}$:
\begin{eqnarray}
{\cal U} ={1\over \sqrt{2}}\pmatrix{1 & -i &
0\cr 0 & 0 & -\sqrt{2}\cr -1 & -i & 0\cr}
\end{eqnarray}}
in the following way:
\begin{eqnarray}
{\bf J}_1 = {1\over \sqrt{2}}\pmatrix{0 & 1 & 0\cr
1 & 0 & 1\cr
0 & 1 & 0\cr},\quad
{\bf J}_2 = {i\over \sqrt{2}}\pmatrix{0 & -1 & 0\cr
1 & 0 & -1\cr
0 & 1 & 0\cr}, \quad
{\bf J}_3 = \pmatrix{1 & 0 & 0\cr
0 & 0 & 0\cr
0 & 0 & -1\cr}\quad.
\end{eqnarray}
The eigenvalues of the operator ${\bf J}\cdot {\bf n}$ could be
$h=\pm 1, \, 0$.
As opposed to a spin-1/2 case one has $9=3^2$ linear independent
matrices forming the complete set. They can be chosen
from the following set of the ten symmetric matrices
\begin{eqnarray}
J_{00} &=& \openone,\quad J_{0i} = J_{i0} = {\bf J}_i\quad,\\
J_{ij} &=& {\bf J}_i {\bf J}_j + {\bf J}_j {\bf J}_i - \delta_{ij}\quad.
\end{eqnarray}
The condition $J_{\mu\mu} =0$ eliminates one of $J_{\mu\nu}$
matrix ({\it e.~g.}, $J_{00}$). Following to main points of
the preceding discussion
let me consider relations between left- and right- spinors.  The
following form of the Ryder-Burgard relation:
\begin{equation}
\phi_R^{\pm, 0} (\overcirc{p}^\mu) = e^{i\alpha_{\pm,0}} \phi_L^{\pm, 0}
(\overcirc{p}^\mu)\quad,\quad
\phi_L^{\pm, 0} (\overcirc{p}^\mu) = e^{-i\alpha_{\pm,0}} \phi_R^{\pm, 0}
(\overcirc{p}^\mu)
\end{equation}
is very similar to the first
form of the relation in a spin-1/2 case.  In the process of deriving this
relation we used that any tensor can be expanded in a direct product of
two vectors. The equation obtained by using the Wigner postulate (item 1,
$m^2\neq 0$)
\begin{equation}
\left [\gamma_{\mu\nu} p^\mu p^\nu - m^2 {\cal T}\right ] \Psi (p^\mu) = 0
\end{equation}
in the case $\alpha_{\pm, 0} =0$ coincides with the Weinberg equation
and, after taking into account $\alpha_{\pm, 0}=\pm \pi$, with the modified
equation obtained by Ahluwalia~\cite{DVA0} in the framework of
the FNBWW-type quantum field theory~\cite{Nigam,Wigner2}.

As for the second form (connecting $\phi_{L,R}$ and $\phi_{L,R}^*$)
one has an essential difference from the spin-1/2 consideration.
Expanding the ${\cal B}$ matrix, $\phi_R (\overcirc{p}^\mu)
= {\cal B}\phi_L^*(\overcirc{p}^\mu)$,
in the other complete set,\footnote{The explicit form
of the Wigner operator $\Theta_{[1]}$ for spin 1
has been given in refs.~\cite{DVAG,Dowker} in the isotropic basis:
\begin{eqnarray}
\Theta_{[1]} = \pmatrix{0 & 0 & 1\cr
0 & -1 & 0\cr
1 & 0 & 0\cr}\quad.
\end{eqnarray}}
namely, $J_{\mu\nu}\Theta_{[1]}$ we come  to
\begin{equation}
\phi_R^{\pm, 0} (\overcirc{p}^\mu)
\,=\, e^{i\,\beta_{\mp, 0}} \,\Theta_{[1]} \,[\phi_L^{\pm, 0}
(\overcirc{p}^\mu)]^* \quad,\quad \phi_L^{\pm, 0} (\overcirc{p}^\mu) \,=
\,e^{i\,\beta_{\mp, 0}} \,\Theta_{[1]} \,
[\phi_R^{\pm, 0} (\overcirc{p}^\mu)]^*
\quad.
\end{equation}
This fact is connected with another property of the Wigner operator:
$\Theta_{[j]} \Theta_{[j]} = (- 1)^{2j}$. As a result, we obtain
\begin{eqnarray}
\Psi (p^\mu) = \pmatrix{\phi_R (p^\mu)\cr
\phi_L (p^\mu)\cr} =
e^{i\,\beta_{\pm, 0}} \,\pmatrix{0 & \Theta_{[1]} {\cal K}\cr
\Theta_{[1]} {\cal K} & 0\cr} \pmatrix{\phi_R (p^\mu)\cr
\phi_L (p^\mu)\cr} = \Gamma_5 S^c_{[1]} \Psi (p^\mu)\quad,
\end{eqnarray}
provided that the charge conjugation operator $S^c_{[1]}$ is chosen
like ref.~\cite{DVA0,DVAG} in accordance with the FNBWW construct.

Next, let me draw your kind attention to other possibilities
of description of arbitrary spin particles.
In general, it is possible to choose other representation of the Lorentz
group for describing higher spin particles (see ref.~[6c]).
It is interesting to note that the well-known the Dirac-Fierz-Pauli
equation for any spin has been re-written
in ref.~\cite{Dowker} (see also my recent work~\cite{DDO})
to the form very similar to the spin-1/2 case:
\begin{mathletters}
\begin{eqnarray}
\alpha^\mu \,\partial_\mu \Phi &=& +\,m\Upsilon\quad,\\
\bar\alpha^\mu \,\partial_\mu \Upsilon &=& -\,m \Phi\quad,
\end{eqnarray}
\end{mathletters}
where $\bar \alpha^\mu =\alpha_\mu$ are the matrices that satisfy
all the algebraic relations that the Pauli $2\times 2$
matrices $\sigma^\mu$ do, except for completeness.
The object $\Phi$ belongs to the $(j,0)\oplus
(j-1,0)$ representation of the Lorentz group and the $\Upsilon$, to the
$(j-1/2,1/2)$ representation.
Is there exist the analog of the Ryder-Burgard relation which could
be proposed in the framework of the
Dirac-Fierz-Dowker construct for any spin?

Finally, in my presentation the attempt was undertaken to understand,
how all possible relations between
basis vectors of the different representation
space ({\it e.g.}, between right- $\phi_R (\overcirc{p}^\mu)$
and left-handed spinors $\phi_L (\overcirc{p}^\mu)$ that are known
to become interchanged under parity conjugation~\cite{Ryder})
define dynamical equations. It was found that from
a mathematical viewpoint  the well-known  equations are
the particular cases only. The analysis reveals
that the choice of spinorial basis in $(j,0)\oplus (0,j)$
representation space has profound significance
for dynamical evolution of the physical systems.

\acknowledgments
I appreciate
the help of Prof. D. V. Ahluwalia whose articles and preprints
are the ground for this work. The valuable advises of Prof. A. F. Pashkov
were helpful in realising the ideas presented here and in my previous
publications. I thank my students assisting at the course of advanced
quantum mechanics for asking many good questions.

I am grateful to Zacatecas University for a professorship.


\begin{references}
\footnotesize{
\bibitem{DFT} V. V. Dvoeglazov, Yu. N. Tyukhtyaev and R. N. Faustov,
Mod. Phys. Lett. A{\bf 8} (1993) 3263;
Fiz. Elem. Chast. At. Yadra {\bf 25} (1994) 144
[English translation: Phys. Part. Nucl. {\bf 25} (1994) 58]

\bibitem{Rosner} J. L. Rosner, {\it Overview of the Standard Model.}
Preprint EFI 94-59 (hep-ph/9411396), Nov. 1994, to be published
in {\it Proc. of the Joint US-Polish Workshop on Physics
from Planck Scale to Electroweak Scale. Warsaw, Poland, 21-24 Sep., 1994}

\bibitem{QCD} R. K. Ellis, {\it Status of QCD.} Preprint
FERMILAB-CONF-93/011-T, Jan. 1993, presented at {\it the 7th Meeting
of the APS Division of Part. and Fields, Fermilab, Nov. 10-14, 1992}

\bibitem{Furry} W. H. Furry, Phys. Rev. {\bf 54} (1938) 56;
ibid {\bf 56} (1939) 1184

\bibitem{Dirac} Not all physicists are so enthusiastic in evaluations
of the status of modern physics theories. For
instance, let me remind what claimed P. A. M. Dirac in his last works.
{\it E.~g.}, in [{\it Mathematical Foundations of Quantum Theory.}
(Academic Press, Inc., 1978), p. 1] he doubts mathematical
grounds of modern physics: ``Any physical or phylosophical
ideas that one has must be adjusted to fit the mathematics.
Not the other way around.
Too many physicists are inclined to start from some preconceived
physical ideas and then to try to develop them and find a mathematical
scheme that incorporate them. Such a line of attack is unlikely
to lead to success...

\,\,\,\,\,The appearance of this [Dirac] equation did not solve
the general problem of making quantum
mechanics relativistic. It applied only to the problem of a single
electron, not several particles in interaction... When one tried to
solve it, one always obtained divergent integrals...
Rules for discarding the infinities [(renormalization) have
been developed]. Most physicists are very satisfied with
this situation. They argue that if one has rules for doing calculations
and the results agree with observation, that is all that one requires.
But it is not all that one requires. One requires a single comprehensive
theory applying to all physical phenomena. Not one theory for
dealing with non-relativistic effects and a separate disjoint theory
for dealing with certain relativistic effects.
Furthermore, the theory has to be based on sound mathematics, in
which one neglects only quantities that are small. One is not
allowed to neglect infinitely large quantities. The renormalization
idea would be sensible only if it was applied with finite renormalization
factors, not infinite ones.
For these reasons I find the present quantum electrodynamics quite
unsatisfactory. One ought not to be complacent about its faults.
The agreement with observation is presumably a coincidence, \, just
like the original calculation of the hydrogen spectrum with
Bohr orbits. Such coincidences are no reason for turning a blind eye to
the faults of a theory.
Quantum electrodynamics ... was built up from physical ideas that were not
correctly incorporated into the theory and it has no sound mathematical
foundation. One must seek a new relativistic quantum mechanics and
one's prime concern must be to base it on sound mathematics."

\bibitem{Weinberg} S. Weinberg, Phys. Rev. B{\bf 133} (1964),
1318;  ibid {\bf 134} (1964) 882; ibid {\bf 181} (1969) 1893.
In these papers the pioneer study of the $(j,0)\oplus (0,j)$
representation space has been undertaken. In fact, this way of
description of particles of arbitrary spin originates from
the classical work of E.~Wigner, ref.~\cite{Wigner}. Unfortunately,
one of remarkable statements of the second paper of S. Weinberg
was not realised before an appearance of the paper~\cite{DVA00}. It
deals with the massless
first-order equations (4.19,4.20) for arbitrary spin
(as a particular case, with the first-order equations
(4.21,4.22) for spin $j=1$). ``The fact that these field equations
are of first order for any spin seems to me to be of no great
significance... "(II, p. B888). ``A field with $2j+1$ components,
which are constructed as a linear combination of $2j+1$ independent
creation and/or annihilation operators, can satisfy only
the trivial wave equations $(\Box^2 - m^2)^N \psi =0$"
(III, p. 1896).

{\it Cf.} with the thoughts of P. A. M.
Dirac~\cite{Dirac}.

\bibitem{Wigner} E. P. Wigner, Ann. Math. {\bf 40} (1939) 149

\bibitem{DVA00} D. V. Ahluwalia and D. J. Ernst, Mod. Phys. Lett.
A{\bf 7} (1992) 1967

\bibitem{DVO-old} V. V. Dvoeglazov, Rev. Mex. Fis. Suppl. {\bf 40} (1994)
352

\bibitem{Ryder} L. H. Ryder, {\it ``Quantum Field Theory"} (Cambridge
University Press, Cambridge, UK, 1987), \S 2.3

\bibitem{DVA0} D. V. Ahluwalia, M. B. Johnson and T. Goldman,
Phys. Lett.  B{\bf 316} (1993) 102;
D. V. Ahluwalia and T. Goldman, Mod. Phys. Lett.  A{\bf 8} (1993)
2623

\bibitem{DVAG} D. V. Ahluwalia, {\it ``Incompatibility of Self-Charge
Conjugation with Helicity Eigenstates and Gauge Interactions."} Preprint
LA-UR-94-1252, Los Alamos, Apr. 1994; {\it ``McLennan-Case Construct for
Neutrino, its Generalization, and a Fundamentally New Wave Equation."}
Preprint LA-UR-94-3118, Los Alamos, Sept. 1994

\bibitem{Wigner2} E. P. Wigner, in {\it ``Group theoretical concepts
and methods in elementary particle physics -- Lectures of the Istanbul
Summer School of Theoretical Physics, 1962".}
Ed. F. G\"ursey

\bibitem{Sachs} T. Ohmura (Kikuta), Progr. Theor. Phys. {\bf 16}
(1956) 684; M. Sachs, Ann. Phys. {\bf 6} (1959) 244;
M. Sachs and S. L. Schwebel, J. Math. Phys. {\bf 3}
(1962) 843

\bibitem{Itzykson} C. Itzykson and J.-B. Zuber, {\it ``Quantum
field theory."} (McGraw-Hill Book Co., 1980)

\bibitem{Novozh} Yu. V. Novozhilov, {\it ``Vvedenie v teoriyu
elementarnykh chastitz."} (Nauka, Moscow, 1971) [English translation: {\it
``Introduction to Elementary Particle Theory."} (Pergamon Press, Oxford, UK,
1975), p. 96]

\bibitem{Nigam} B. P. Nigam and L. L. Foldy, Phys. Rev. {\bf 102}
(1956) 1410

\bibitem{Sokolik} G. A. Sokolik, ZhETF {\bf 33} (1957) 1515 [English
translation: Sov. Phys. JETP {\bf 6} (1958) 1170]

\bibitem{Cartan} \'E. Cartan, {\it ``Lecons sur la Th\'eorie des
spineurs."} (Hermann, Paris, 1938)

\bibitem{Gelfand} I. M. Gelfand and M. L. Tsetlin, ZhETF {\bf 31}
(1956) 1107 [English translation: Sov. Phys. JETP {\bf 4} (1957) 947]

\bibitem{Ramond} P. Ramond, {\it Field Theory: A Modern Primer.}
(Addison-Wesley Pub. Co., USA, 1989)

\bibitem{Majorana} E. Majorana, Nuovo Cim. {\bf 14} (1937) 171
[English translation: Tech. Trans. TT-542, Nat. Res. Council
of Canada]

\bibitem{Case}  K. M. Case,  Phys. Rev. {\bf 107} (1957) 307

\bibitem{DVA} D. V. Ahluwalia, M. B. Johnson and T. Goldman,
Mod. Phys. Lett.  A{\bf 9} (1994) 439

\bibitem{DV95} V. V. Dvoeglazov, {\it ``A note on the Majorana
theory for $j=1/2$ and $j=1$ particle states."} Preprint
EFUAZ FT-94-10, Dec. 1994, reported at the XVIII Oaxtepec Nuclear
Physics Symp., Jan. 4-7, 1995

\bibitem{Var} D. A. Varshalovich, A. N. Moskalev and V. K.
Khersonski\u{\i}, {\it Kvantovaya teoriya uglovogo momenta.}
(Nauka, Leningrad, 1975) [English translation: {\it Quantum
Theory of Angular Momentum.} (World Sci. Pub., Singapore, 1988)]

\bibitem{Dowker} J. S. Dowker and Y. P. Dowker, Proc. Roy. Soc. A{\bf 294}
(1966) 175; J. S. Dowker, ibid {\bf 297} (1967) 351

\bibitem{DDO} V. V. Dvoeglazov, Nuovo Cim. {\bf 107}A (1994) 1758
}
\end{references}
\end{document}